\documentclass[conference]{IEEEtran}
\IEEEoverridecommandlockouts
\usepackage{cite}
\usepackage{amsmath,amssymb,amsfonts}
\usepackage{algorithm}
\usepackage{caption}
\usepackage{subcaption}
\usepackage{graphicx}
\usepackage{textcomp}
\usepackage{xcolor}
\usepackage{arydshln}
\usepackage{tikz}
\usetikzlibrary{patterns}

\usepackage{comment}
\usepackage{todonotes}

\usepackage{textpos}

\usepackage{algorithm}
\usepackage{algpseudocode}
\def\BibTeX{{\rm B\kern-.05em{\sc i\kern-.025em b}\kern-.08em
    T\kern-.1667em\lower.7ex\hbox{E}\kern-.125emX}}

\usepackage{verbatim,fancyheadings,bbm}

\begin{document}

\title{A Local Machine Learning Approach for Fingerprint-based Indoor Localization\\
\thanks{N. Agah and B. L. Evans were supported by NVIDIA, an affiliate of the 6G@UT Research Center within the Wireless Networking and Communications Group at The University of Texas at Austin.}
}

\author{
    \IEEEauthorblockN{Nora Agah\IEEEauthorrefmark{1}, Brian Evans\IEEEauthorrefmark{2}, Xiao Meng\IEEEauthorrefmark{3} and Haiqing Xu\IEEEauthorrefmark{4}}
    \IEEEauthorblockA{\IEEEauthorrefmark{1}\IEEEauthorrefmark{2}\textit{Dept. of Electrical and Computer Engineering, The University of Texas at Austin, Austin, TX USA}}
    \IEEEauthorblockA{\IEEEauthorrefmark{3}\textit{McCoy College of Business Administration, Texas State University, San Marcos, TX USA}}
    \IEEEauthorblockA{\IEEEauthorrefmark{4}\textit{Dept. of Economics, The University of Texas at Austin, Austin, TX USA}}   
    \IEEEauthorrefmark{1}norakagah@utexas.edu, \IEEEauthorrefmark{2}bevans@ece.utexas.edu, \IEEEauthorrefmark{3}mengxiao23@gmail.com, \IEEEauthorrefmark{4}h.xu@austin.utexas.edu
}

\maketitle

\begin{textblock*}{210mm}(-1.5cm,-7.5cm)
©2023 IEEE.  Personal use of this material is permitted.  Permission from IEEE must be obtained for all other uses, in any current or future media, including reprinting/republishing this material for advertising or promotional purposes, creating new collective works, for resale or redistribution to servers or lists, or reuse of any copyrighted component of this work in other works.
\end{textblock*}

\thispagestyle{plain}
\pagestyle{plain}

\begin{abstract}
Machine learning (ML) solutions to indoor localization problems have become popular in recent years due to high positioning accuracy and low cost of implementation.  This paper proposes  a novel local nonparametric approach for solving localizations from high-dimensional Received Signal Strength Indicator (RSSI) values. Our approach  consists of a sequence of classification algorithms that sequentially narrows down the possible space for location solutions into smaller neighborhoods. The idea of this sequential  classification  method is similar to the decision tree algorithm, but a key difference is our splitting of the dataset at each node is not based on  features of input (i.e. RSSI values), but some  discrete-valued variables generated from the output variable (i.e.  the 3D real--world coordinates). The strength of our localization solution  can be tuned to problem specifics by the appropriate choice of how to sequentially partition the the space of location into smaller neighborhoods.   Using the publicly available indoor localization dataset {\it UJIIndoorLoc}, we evaluate our proposed method vs. the global ML algorithms for the dataset. The primary contribution of this paper is to introduce a novel local ML solution for indoor localization problems.
\end{abstract}


\section{Introduction}
\subsection{Motivation}

%
%
%

With the development of wireless access infrastructure and the popularity of mobile devices, indoor-location-based services, i.e., finding the position of a person in indoor environments, have become essential in many applications. Different from the Global Positioning System (GPS), the fingerprint-based indoor positioning technology uses received WiFi signal strengths, i.e. {\it Received Signal Strength Indicator} (RSSI), from ubiquitous wireless access points (AP). It has become a suitable substitution solution  to localization problems in indoor environments as GPS signals cannot penetrate well \cite{b1}.

\begin{figure}[tbhp]
\centerline{\includegraphics[scale=0.5]{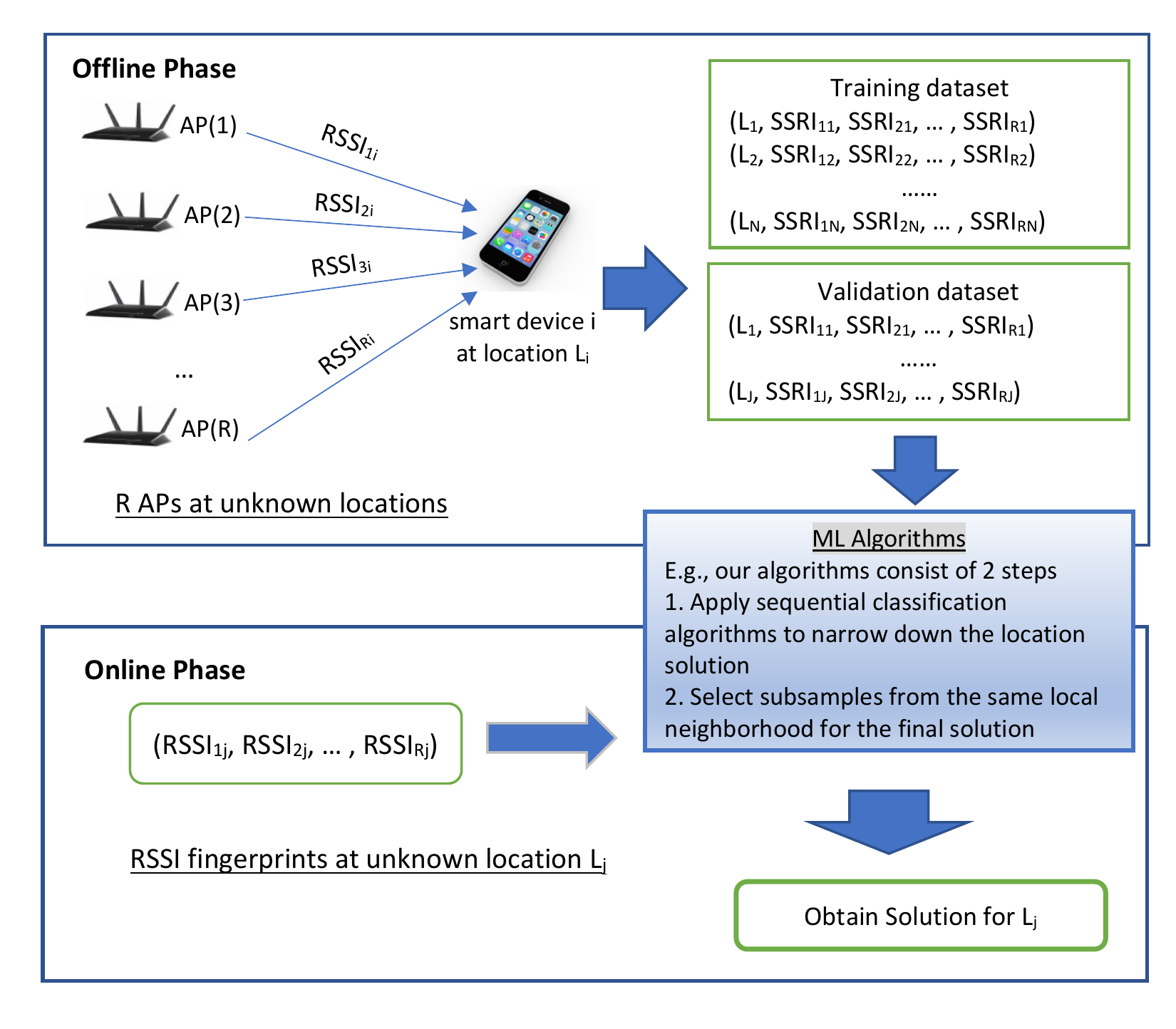}}
\caption{Typical fingerprint positioning system with an offline phase and an online phase.}
\label{fig1}
\end{figure}

Machine Learning (ML) is one of the most promising methods for solving fingerprint-based indoor localization problems due to its high accuracy and simplicity. Because of the interference of many noise factors in practical indoor environments, it is difficult to build an  accurate theoretic model for the wireless propagation that describes the actual relationship between the real-world locations of mobile devices and the signal strengths received by various AP. To overcome this, a variety of ML approaches  using large-sized big data  have been introduced to solve fingerprint-based indoor localization problems; See e.g. \cite{b3,b2} for a detailed review on this literature. 

ML methods for fingerprint-based indoor positioning, e.g. \cite{b5,b6,b7,b8,b4}, usually include two main phases shown in Fig. \ref{fig1}. Specifically, the offline phase collects fingerprint observations, split into training and validation datasets. An algorithm uses these two datasets jointly to train the mapping  between RSSI readings and position coordinates. In the online phase, the algorithm uses the obtained mapping to localize each location requester from its observed RSSI readings. Typical ML algorithms, often providing general purpose solutions, usually establish the relationship between 3D position coordinates and large dimensional RSSI readings by using global parametric estimates,
 equipped with model selection techniques that determine which parametric model/features to be chosen.

%

While ML techniques provide effective solutions to indoor positioning, there are still challenges arising  due to using observational data collected for indoor localization practice; see e.g. \cite{b3,b2}. First, there is two-sided heterogeneity regarding the signal strength measurements (i.e. RSSI) due to dissimilar smart devices as well as heterogeneous environments or configurations of APs. To see this,  consider a smart device at some fixed location, for which we obtain a number of RSSI measurements from  various APs at different places. The signal strength  need not simply reflect the distances of these APs to the smart device, since the configuration of APs as well as obstacles in their local environments also play a major role in the actual wireless propagation process. On the other hand, the same AP could record quite different RSSI readings sent by different smart devices located at even the same location \cite{b12}. Because of such heterogeneity, it is difficult to use a global parametric model to capture  all the relationship between RSSI measurements and locations.  Some RSSI readings  in the dataset might be highly informative on whether a smart device is located in a specified building/area or not, but it may not be  further useful to tell the  exact location of the device within the building/area. It's unclear whether these features would even be selected in an ML model.

%
%
%
%
%
%
%

Second, the indoor localization problems  involve large dimensional RSSI readings, measured by  APs deployed at various locations.  Intuitively, increasing the number of APs should improve the accuracy of localization solutions. However, the relationship between localization solutions and these RSSI readings will be more complicated, and the complexity escalates quickly as the size of the indoor localization problem (i.e. the space of possible locations under consideration and the dimensionality of RSSI measurements) expands.

To address these difficulties, we propose a new ML solution to indoor localization problems, which is based on the local approach in the nonparametric estimation literature. In the literature, the local method, e.g. Nadaraya–Watson kernel regression,  uses a kernel  function to obtain a locally weighted average estimator for the expectation of the output conditional on some input value. However, the kernel regression suffers from the curse of dimensionality since the dimensionality of the input here, i.e. RSSI measurements, is large. That being said, it is not realistic to directly control RSSI measurements within a local neighborhood of some given RSSI value to estimate its location.  Instead, we introduce some binary-valued (or discrete) features of the output variable, which can be easily constructed from the observed real-word coordinates in the sample observations. For instance, a feature could be whether an observation from a specific area/building or not.  Each feature partitions the space of possible locations into two and will be learned by a binary classification algorithm  at nearly 100\% accuracy. Thus, the relationship between  the high-dimensional RSSI readings and these binary features is approximately deterministic. Moreover, these binary classification algorithms are implemented sequentially, structured as a tree diagram, which narrows down the location solution to a smaller region in each step.  The binary classification results are used to select a subsample of observations located within the same region  for the next stage training.  

Our method is motivated by the common sense that in a causal inference model, it is the location of mobile devices to determine these RSSI readings, rather than the other way around.  Moreover, it might be possible to learn some features of the location solution at nearly perfect accuracy, without being affected by the heterogeneity in observations.   Therefore, the proposed local method aims at using observations in the training dataset that are close  to the (unknown) location solution to learn how the exact location depends on the  RSSI readings from mobile devices at  a local neighborhood.  By shrinking the space of possible location solutions, the relationship between RSSI readings and signal locations can be greatly simplified, which thereafter improves the accuracy of ML localization solutions.  We evaluate our method by using publicly available data provided by the {\it UJIIndoorLoc} database, as used for the EvAAL 2015 competition on indoor localization \cite{b11}.

One fundamental idea in ML is the so-called bias-variance tradeoff. By narrowing down the location solution to smaller neighborhoods, this helps reduce the bias of the localization solution at the cost of the variance, since we are selecting a subsample of smaller size for the final localization solution. In particular, we specify a stopping rule for the sequential classification procedure, which is satisfied if the subsample selected by a further partition/feature of the current location space is not sufficiently large, or there does not exist any partition that can be learned accurately (e.g. $\geqslant 98\%$). With the stopping rule satisfied, we further implement a final-stage ML algorithm to solve the final location. 

The rest of this paper is organized as follows: in Section II, we start with describing the {\it UJIndoorLoc} dataset, following with a brief literature review, and then introducing our method. Next, we show the localization results using our method for the UJIIndoorLoc validation dataset and compare them to the competing teams of the EvAAL 2015 competition. Our conclusions and discussions are provided in Section IV. 







\section{UJIIndoorLoc Dataset, Model and Methods}
This section describes in detail the dataset, model and the proposed structural ML approach to the indoor localization problem. A brief literature review of the related ML work on indoor localization problems is also provided.

\subsection{Description of UJIIndoorLoc Dataset}
To study the indoor localization problem, we employ {\it UJIIndoorLoc}, the biggest open-access  database in the indoor localization literature. This dataset contains a training set  of 19,937 observations, and a validation set of  1,111 observations, collected by 18 different users with 25 different mobile devices. In the positioning environment, there are 520 Wireless APs distributed within the three buildings (i.e. Building 0, 1, and 2) with four to five floors. A picture of the outdoor environment with three buildings is shown in Fig. \ref{fig2:first}. In the dataset, each observation provides the real-world coordinates (i.e.  Longitude, Latitude, and floor of the building) of a mobile device\footnote{Besides the real-word coordinates, the {\it UJIIndoorLoc} dataset also contains information on Space ID, User ID, Phone ID, and Timestamp. }, and $520$ RSSI values of all the Wireless APs detected by the mobile device. 

The signal strength measure is the RSSI, where $-100$ dBm is equivalent to a weakest signal, whereas $0$ dBM means that the detected AP has an extremely good signal. In addition, if an AP's signal is not received by the mobile device,  we code it by $-105$ dBm.

For simplicity, we denote $N=19,937$ and $R=520$ as  the training sample size and the number of APs/features, respectively. Moreover, let $S_i\equiv(s_{i1},\cdots,s_{iR})$ be $R$-dimensional real vector of the RSSI measurements in the $i$-th observation and  $L_i\equiv (\ell_{ai},\ell_{oi},\ell_{fi})$ be the location information of the mobile device, denoting Longitude, Latitude, and Floor, respectively.\footnote{Note that Building ID, a location variable also provided by the dataset, is fully determined by the Longitude and Latitude of the location.}  Hence, the $i$-th  observation in the (training or testing) dataset can be described as $(S_i,L_i)$. Moreover, we denote the training dataset and the validation dataset by $\mathcal T$ and $\mathcal V$, respectively. Furthermore, we denote  $\mathcal L$ as the space of all real-world coordinates under our consideration, i.e.  $\mathcal L$ consists of all the possible coordinates of three multi-floor buildings.


\subsection{Related work}
Recently, deep learning has been introduced for RSSI-based indoor localization solutions; see e.g. \cite{b8,b4,b9,b10} among many others. Their approaches take high-dimensional RSSI readings $S_i$ as a direct input and use different neural networks for location solutions as the output.  Alternatively, the k-nearest neighbors (KNN) approach directly maps RSSI readings to locations by detecting the most similar fingerprints in the sample and applying majority rules to estimate the location solution. For instance, \cite{b7} proposes a KNN localization solution which sequentially determines the Building ID, Floor, and then the exact longitude and latitude as a filtering process. \cite{b5} also suggests a KNN algorithm to select most similar fingerprints, but their location solution is based on the Maximum Likelihood Estimation (MLE). Moreover, \cite{b6} propose a two-stage calibrated weighted centroid localization algorithm which takes a flavor of structural analysis. In the first step, they estimate the virtual positions of the APs by using the weighted centroid algorithm which does not necessarily need to closely match the real positions of the APs. Next, they calculate positions from observed RSSI readings  by another weighted centroid algorithm, i.e. calculating the weighted  sum of the estimated AP's virtual positions.

\subsection{Proposed Model and Learning Methods}
In this subsection, we introduce a new ML approach that conducts a sequence of binary classification ML algorithms to gradually pin down a localization solution.  We also provide insight on how the proposed method is related to the traditional ML approach, but effectively incorporates the common sense and domain knowledge for localization solutions.

  Consider the following general nonparametric model for the RSSI readings: for observation $i=1,\cdots,N$, there is 
\[
S_{i}=m(L_i, \epsilon_i,\eta), 
\]where $S_i\in\mathbb R^{R}$ is $R$-dimensional RSSI measurements, $L_i\equiv (\ell_{ai},\ell_{oi},\ell_{fi})$ and $\epsilon\in\mathbb R^{d_\epsilon}$ is the location and unobserved features, respectively, of  mobile device in the $i$--th observation, and $\eta\in\mathbb R^{R}$ is the error term of the model. Moreover, vector-valued function $m$ describes the structural relationship between the features (observed and also unobserved) of a mobile device and $R$-dimensional RSSI readings. In addition, we use $m_r$ to denote the $r$--th component of $m$, which links the RSSI measurement from the $r$--th AP to its structural inputs, i.e. 
\[
S_{ir}=m_r(L_i, \epsilon_i,\eta_r).
\]Furthermore, if provided with the marginal distribution of the structural inputs $(L_i,\epsilon_i,\eta_r)$, one could derive $\mathbb E(L_i|S_i)$ under the Bayes' rule.  We denote $h_0(\cdot)\equiv \mathbb E(L_i|S_i=\cdot)$ as an infinite-dimensional object of interest to be estimated.  Most of the traditional ML localization solutions are to estimate $h_0(\cdot)$ directly from a large-sized training sample, from which estimates  $\hat h(S_i)$ will serve as a  solution for $L_i$.\footnote{In Bayesian methods, an alternative solution is to find ${\arg\max}_{\ell}\mathbb P(L_i=\ell|S_i)$; see \cite{b16}.}  For instance, neural networks are proved to be effective for  constructing  non-linear estimates of $h$.

To introduce our sequential learning algorithm, consider the following simple two-step learning algorithm. First,  let $\{\mathcal L_0, \mathcal L_1\}$ be a binary partition of the location space $\mathcal L$ under consideration, i.e. $\mathcal L_1\cup \mathcal L_0=\mathcal L$ and  $\mathcal L_1\cap \mathcal L_0=\emptyset$, where the partition is implemented by using a simple hyperplane  separating the space $\mathcal L$ into two. Next, we denote a dummy variable $Z_i=\mathbbm 1 (L_i\in \mathcal L_1)$, where $\mathbbm 1 (\cdot)$ is the indicator function. By definition, for $z=0,1$, $Z_i=z$  indicates the location $L_i$ is contained in $\mathcal L_z$. It should be noted that the binary partition is not essential and one could alternatively consider a partition of $\mathcal L$ into $K$ ($K\geq 2$) multiple categories. With the generated feature $Z_i$ of the location $L_i$, we apply a binary classification algorithm to solve $Z_i$ from $S_i$, which is a simpler assignment than the original localization problem. By choosing a proper partition $\{\mathcal L_0,\mathcal L_1\}$, we aim at nearly 100\% accuracy for such a purpose.  Next, define our object of interest  as $h_0^*(S_i,z)\equiv \mathbb E(L_i|S_i,Z_i=z)$ for $z=0,1$, and apply an ML algorithm for estimating $h_0^*(\cdot,z)$.  In particular, to estimate $h_0^*(\cdot,z)$, we use the subsample with $Z_i=z$ in the training dataset, rather than importing all the observations into the algorithm. 


There are two intuitive reasons to consider a localization solution based on the estimates of $h_0^*(\cdot,z)$, rather than $h_0(\cdot)$. First, it is not surprising that the functional relationship of  $h_0^*(\cdot,z)$, i.e. how the expectation of $L_i$ depends on $S_i$ given $Z_i=z$  controlled,\footnote{Note that we can write down $Z_i\approx g(S_i)$ for some binary-valued function $g$, in which the approximation error depends on the accuracy of the binary classification algorithm.}   is conceivably simpler than the relationship of $h_0(\cdot)$. That being said, it is less challenging for a model selection procedure to deal with $h^*_0(\cdot,z)$. As is illustrated below with the {\it UJIIndoorLoc} dataset, as long as we narrow down $L_i$ into a smaller-sized neighborhood,  the performance of ML algorithms is more accurate and more robust, regardless of the choice of tuning parameters. Next, the learning algorithm for $h^*_0(\cdot,z)$ selects a subsample with $Z_i=z$ as inputs, which excludes the sampling noises from the  observations located outside of $\mathcal L_z$. This local approach is particularly powerful in the presence of observational-level  heterogeneity due to mobile devices (not feature-level due to heterogeneous APs). 

In the above algorithm,  a crucial question is: whether to stop any partition under some stopping rule.  The intuition behind  these decisions is similar to the optimal choice of bandwidth  in the kernel estimation  literature, known as  the bias-variance tradeoff.  Using the subsample with $Z_i=z$ reduces the sample size for the estimation of $h_0^*(\cdot,z)$, but the information contained in this subsample is more relevant than the whole sample, thereafter generates less biased estimates. Therefore, we introduce a stopping rule  to prevent small--sized subsamples after controlling for $Z_i=z$.

In addition, we require the binary classification algorithm should achieve nearly perfect accuracy. If there does not exist a binary partition to satisfy this condition, then we should also stop partitioning the location space.  This condition implies that variable $Z_i$, as a binary-valued feature of $L_i$,  should depend on $S_i$ in a deterministic way. Therefore, we could control high-dimensional $S_i$ within some local area by fixing the binary-valued feature variable $Z_i$. The rationality behind our idea is mobile devices close to each other tend to generate similar RSSI measurements.

The above two-step local learning procedure can be extended into a multiple-stage sequential learning procedure that partitions the location space into two smaller regions in each step, until a proper stopping rule is satisfied. 
Algorithm \ref{alg1} provides a decision-tree-type structure of the above sequential  search algorithm. It should be noted that for a given unlabelled leaf $\mathcal L$, we use a hyperplane to partition it into two. Clearly, such kind of binary partitions should not be unique. We may want to start with some ``natural'' partitions or by using unsupervised ML methods, e.g. Spatial clustering. After trying several partitions, we choose the one that achieves the highest accuracy rate among them to split $\mathcal L$.  


\begin{algorithm}[t]
\caption{Sequential classification algorithms}
\begin{algorithmic}[1]
\Statex \textbf{input }Observations $\{(S_{i}, L_i) \}_{i\in \mathcal T}$ and $\{(S_{i}, L_i) \}_{i\in \mathcal V}$
\State Initialize Tree as an unlabeled node, associated with $\mathcal L_0$
\While{there is unlabeled leaf $\mathcal L_{\nu}$}
{\State Split $\mathcal L_{\nu}$  into $\mathcal L_{\nu0}\cup\mathcal L_{\nu1}$ among possible partitions
\State \textbf{for} each partition \textbf{do}
{\State \ \ Navigate observations $L_i\in \mathcal L_{\nu}$ in $\mathcal T$ to leaf $\mathcal L_\nu$
\State \ \ Label leaf observations by: $Z_i=\mathbbm 1(L_i\in\mathcal L_{\nu1})$
\State \ \ Train  $\{(Z_i,S_i): L_i\in\mathcal L_\nu; i\in\mathcal T\}$ by a binary classification ML algorithm
\State \ \ Compute classification accuracy rate $\tau$ by using the validation sample: $\{(Z_i,S_i):L_i\in\mathcal L_\nu; i\in\mathcal V \}$
\State \textbf{end for}
\State Choose one partition  $\mathcal L_{\nu0}\cup\mathcal L_{\nu1}$ that maximizes $\tau$
\If{stopping criterion is satisfied}
{\State Label leaf $\nu$ by $L_\nu$
\Else
\State Put $\mathcal L_{\nu0}$ and $\mathcal L_{\nu1}$ as two unlabeled leaves
\EndIf}
\EndWhile}}
\end{algorithmic}
\label{alg1}
\end{algorithm}

The proposed  algorithm is  a sequential classification neural network procedure, which partitions the location space $\mathcal L$ into a series of subspaces  such that we can classify observations into them sequentially as a decision tree. For instance, given an RSSI readings $S_j$, we could first apply a binary classification algorithm to the whole training sample to determine whether mobile device $j$ is located in Building 0 versus Building 1\&2. If it  classifies $j$'s location into Building 1\&2, we further select the subsample from Building 1\&2 to train a follow-up model that determines whether the mobile device $j$ is located in Building 1 versus Building 2. On the other hand,  if the first classification algorithm classifies  $j$'s location into Building 0 category, we further use a binary classification algorithm to determine whether it's located in the upper level floors (i.e. Floor 2 or 3), or in the lower level floors (i.e. Floor 0 or 1), for which we use subsample of training observations from Building 0. Meanwhile, we use the validation sample to obtain accuracy rates for these binary classification algorithm. We repeat such a partition procedure on the location space   $\mathcal L$ until the stopping rule is satisfied. In each step, we could use multiple different partitions of a space, and then choose the one that achieves highest accuracy.

Although the proposed algorithm is a tree-structured learning procedure, it is different from the commonly used decision tree classification algorithm. In particular, our algorithms build a tree by splitting the outcome space (i.e. the real-world coordinates), rather than input  features (i.e. RSSI readings). Each leaf of the tree is labeled as an area of the whole coordinate space. Therefore, a sequence of binary classification algorithms lead each observation of RSSI readings into a specific leaf and the stopping criterion ensures our tree-structured dynamic classifier  achieves great accuracy by limiting the depth of the tree and  searching optimal  partitions among all possible candidates.

In the last step, we apply a neural network algorithm to a small subsample from the same neighborhood, induced from the above sequential classification algorithm, for  the final localization solution. Specifically, for all the observations  associated with a leaf $\mathcal L_\nu$, we  train a neural network model by selecting a subsample  of observations located within a neighborhood of the leaf. For instance, for all the observations belonging to the lower--level floors (i.e. $\ell_f\leq 1$) of Building 0, we train a neural network model by using all the observations with $\ell_f\leq 2$ from Building 0 in the training dataset as the input.  In this step, the number of algorithms depends on the number of leaves obtained from the above sequential partition procedure. 

\subsection{Clean and Process Raw Data}
To improve the performance of ML algorithms, we need to clean and process the raw data to deal with missing values, noises and  undesirable format. First, we replace the non-detection RSSI value (i.e. $+100$) with $-105$ dBm, which is less than the weakest signal. Next, if an AP's RSSI has no variation within the training and/or validation dataset, we exclude it from our analysis; see e.g. \cite{b11}. Last but not least, we partition the training sample into $m$ (e.g. $m=2$) folders of equal size, according to the time when the fingerprint was collected.  If an AP's location estimates using different folders are quite different from each other, we also exclude the RSSI readings from this AP. This is because the location of this AP might not be fixed during the data collection stage. As a matter of fact, the above noise reduction and filtration reduces the number of RSSI readings $R$ from 520 to $320$, which significantly improves the accuracy as well as the speed of our learning algorithms. 

\section{Performance Metrics and Results}
In this section, we provide the localization results of the proposed method on the {\it UJIIndoorLoc} validation dataset (with 1,111 observations).  Following the rules defined by  the EvAAL 2015 competition on indoor localization, we use the mean error as the metrics for evaluating the results of the proposed method. 

\begin{figure}[htbp]
\centering
\begin{subfigure}[b]{0.2\textwidth}
    \includegraphics[width=1\textwidth]{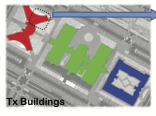}
    \caption{Real-world Environment}
    \label{fig2:first}
\end{subfigure}
\begin{subfigure}[b]{0.22\textwidth}
    \includegraphics[width=1\textwidth]{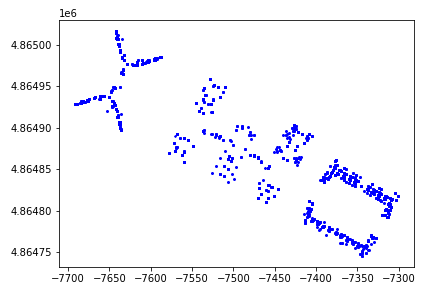}
    \caption{Mobile device locations}
    \label{fig2:second}
\end{subfigure}
\begin{subfigure}[b]{0.22\textwidth}
    \includegraphics[width=1\textwidth]{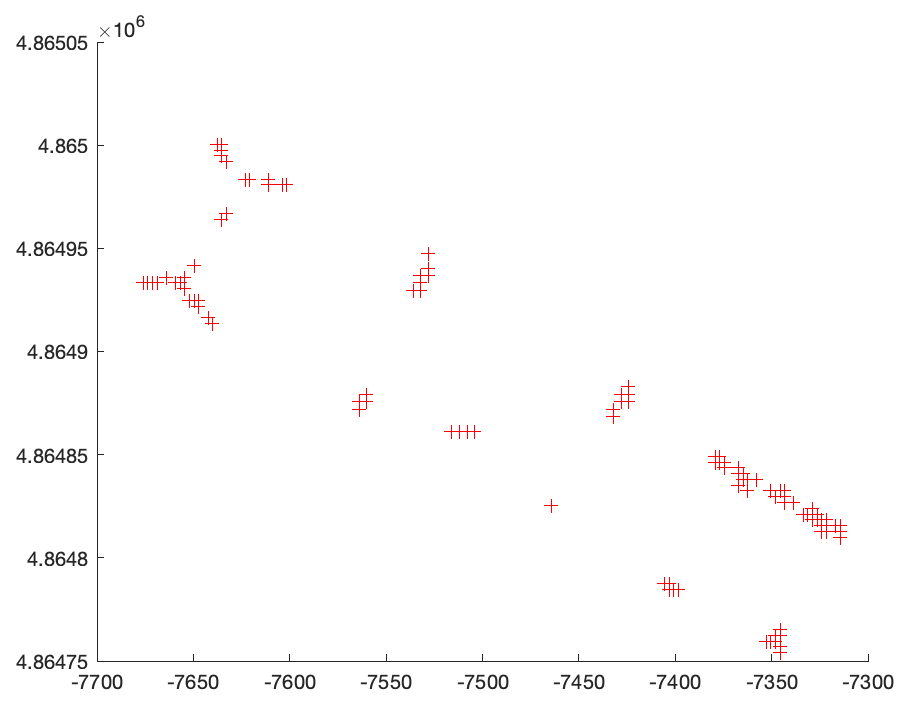}
    \caption{Estimated AP locations}
    \label{fig2:third}
\end{subfigure}
\begin{subfigure}[b]{0.22\textwidth}
    \includegraphics[width=1\textwidth]{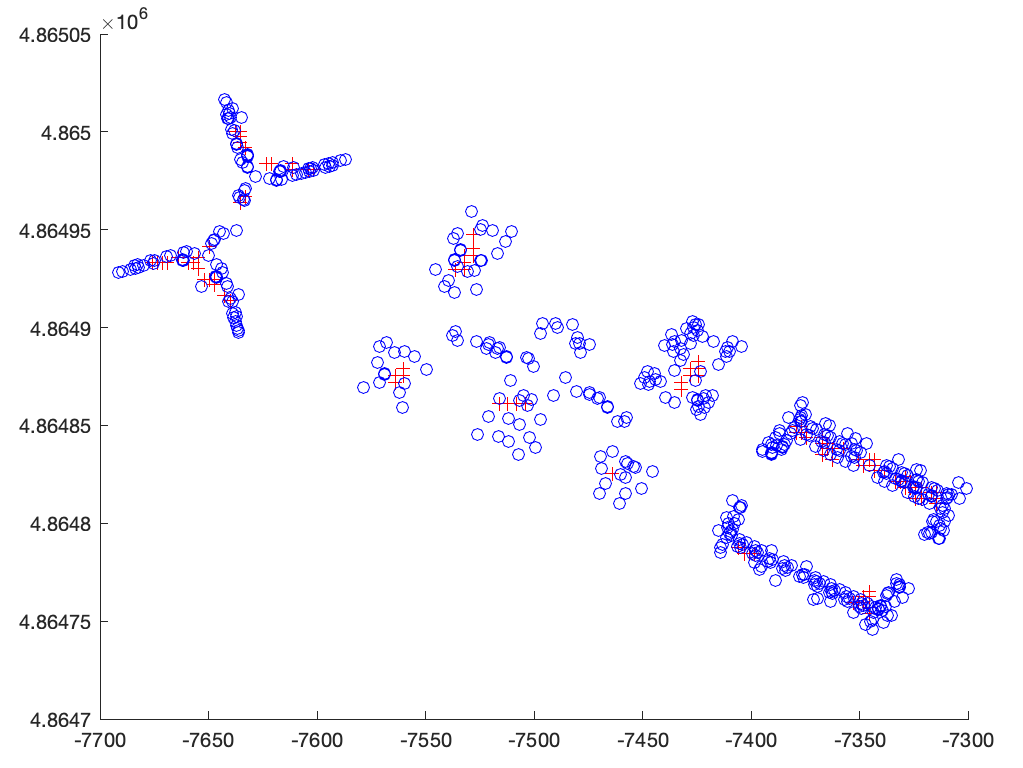}
    \caption{Locations of (b) and (c)}
    \label{fig2:fourth}
\end{subfigure}
        
\caption{Visualization of buildings, phones, and AP locations}
\label{fig2:figures}
\end{figure}

To begin with,  Fig. \ref{fig2:first} and Fig. \ref{fig2:second} respectively shows the  outdoor environment, i.e. the three multi-floor buildings, and the real-world coordinates of mobile devices in the training dataset. In Fig. \ref{fig2:third}, we estimated the locations of all the APs, which provides some intuition for our localization results. Clearly, Building 1 in the middle contains the least number of APs. Therefore, we would expect lower localization accuracy  observations located at Building 1 than those at the other two buildings.  Fig. \ref{fig2:fourth} combines the location information of mobile devices with the estimated locations of APs. Locations of mobile devices evenly spread out everywhere in the tree buildings.

The proposed algorithm first identifies Building ID with a neural network algorithm, achieving 100\% accuracy rate in the this stage, and then learns whether an observation belongs to the upper--level floors (i.e. floor level $\geq 2$) versus the lower--level floors (i.e. floor $0$ or $1$). After these two steps, the stopping rule is applied due to the small number of observations in each neighborhood. Next, we apply neural network algorithms for obtaining solutions of longitude, latitude and floor for all the 1111 observations in the validation dataset. 

 Table \ref{table:2} summarizes  our experiments' results on the floor hit rate and mean positioning error, which shows the proposed algorithm (i.e. Sequential Classification neural network, SCNN) achieves better accuracy than the traditional neural network (TNN) algorithms. For comparison, we also provide results from a two-stage neural network (TSNN) algorithm, which identifies Building ID in the first stage and then applies neural network algorithms to each subsample of training observations according to their Building IDs for localization solutions. It shows that TSNN improves accuracy significantly than the one-stage TNN, but there is still further room for improvements by finer partitioning of the location space (i.e. upper--level  v.s. lower--level floors).

\begin{table}[!h]
\begin{center}
\caption{Localization results on UJIIndoorLoc validation dataset (1111 observations) using neural network algorithms}
\begin{tabular}{lcccccccc}
\hline
       Algorithm  &    Floor hit rate & Mean positioning error (m)\\\hline
TNN &90.64\% & 12.35\\
TSNN &94.48\% &9.81\\
SCNN &95.52\% & 9.68\\
\hline
\end{tabular}
\label{table:2}
\end{center}
\end{table}

Moreover, Fig. \ref{fig1} provides classification accuracy in terms of the floor hit rate of the classification algorithms at each step of SCNN.  In particular, the first step aiming at Building ID achieves 100\% accuracy, while all the three classification algorithms in the second step on identifying upper/lower floor achieve more than 98\% accuracy. For comparison, Fig. \ref{fig1} provides accuracy of each classification algorithm of TSNN. Clearly, additional partitioning of each building into upper/lower floor achieves better accuracy  in every building-specific subsample.

\begin{figure}[tbhp]
\centerline{\includegraphics[scale=0.55]{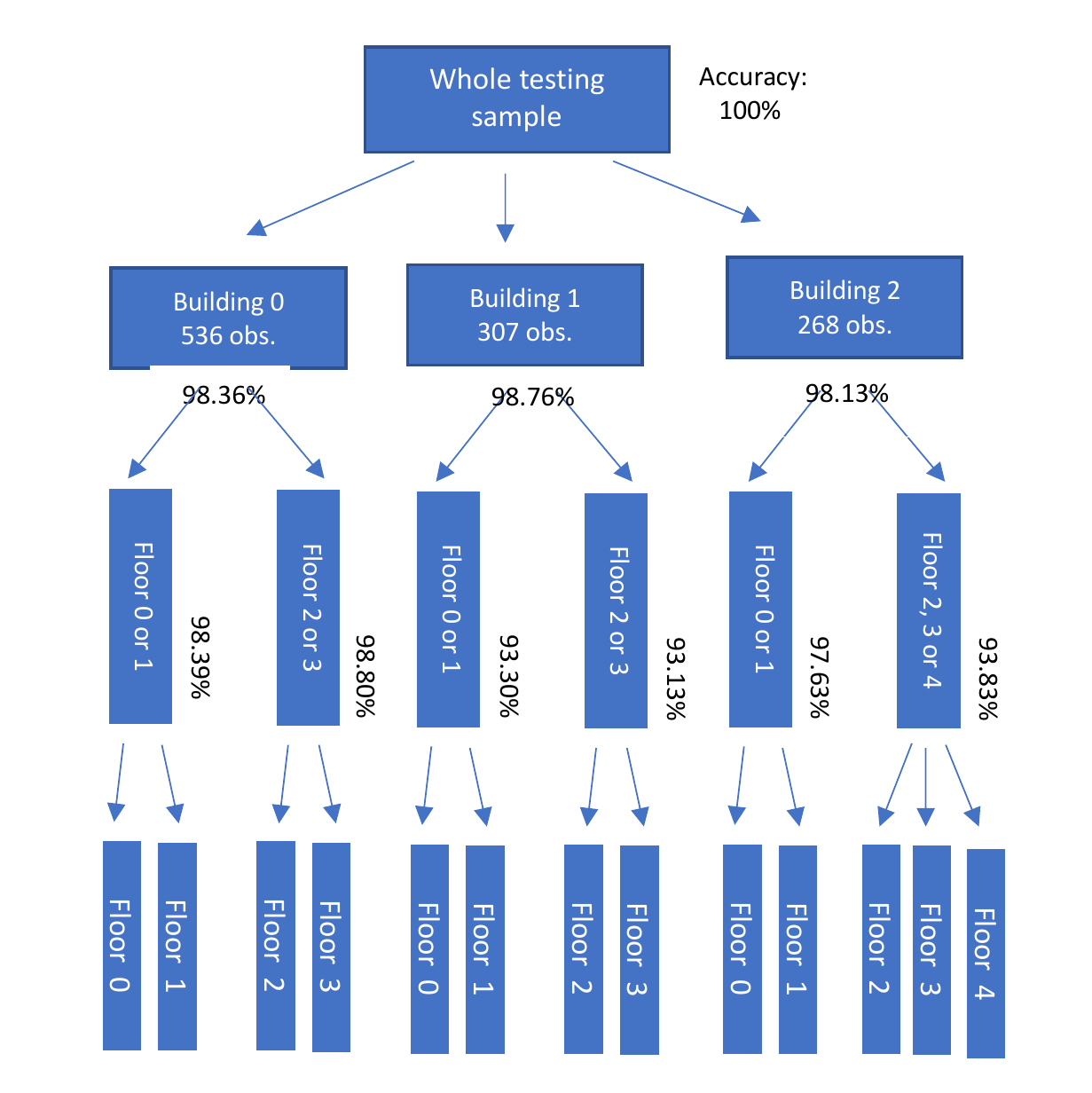}}
\caption{Floor hit rate with two-stage neural network algorithm}
\label{fig}
\end{figure}

\begin{figure}[tbhp]
\centerline{\includegraphics[scale=0.55]{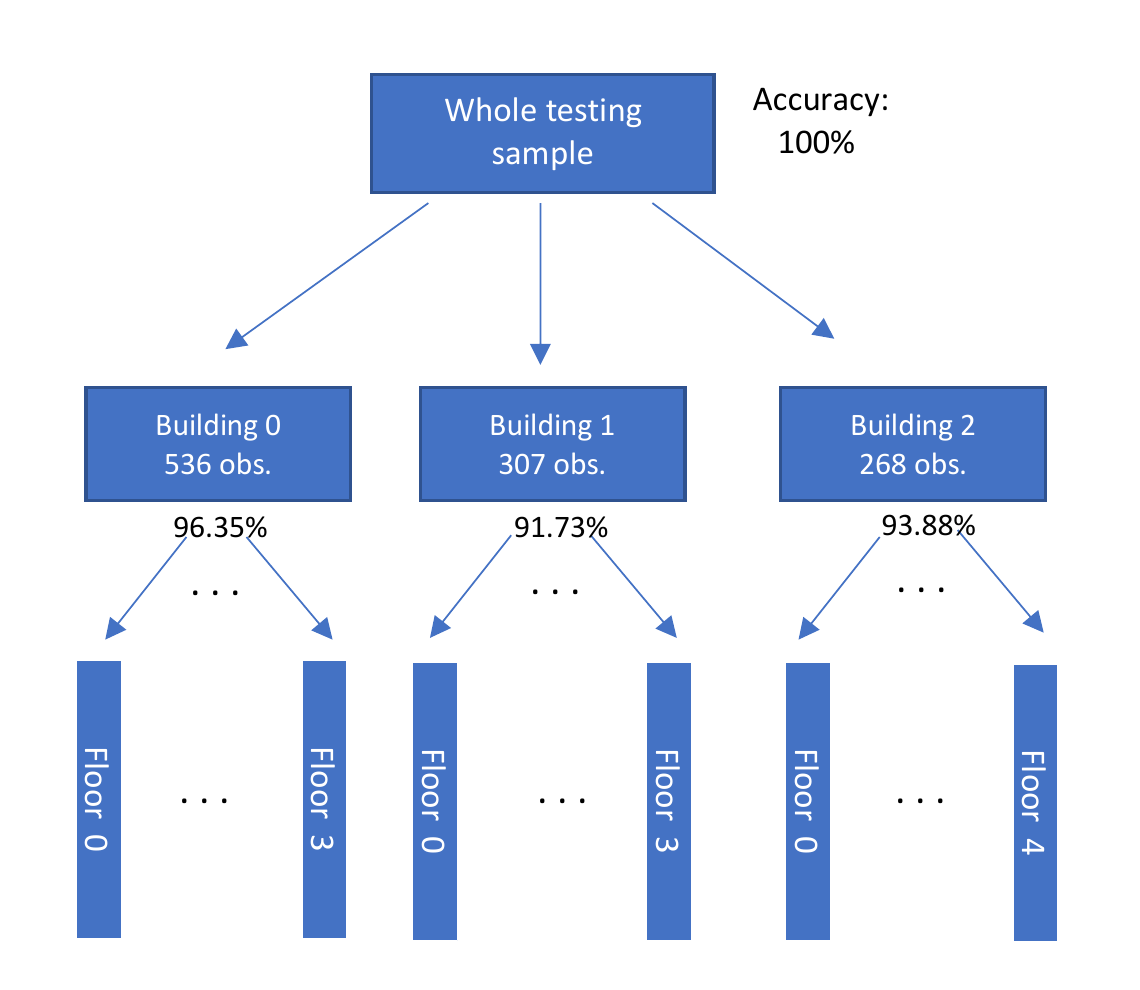}}
\caption{Floor hit rate with two-stage neural network algorithm}
\label{fig3}
\end{figure}

Table \ref{table:3} shows mean positioning error of SCNN for observations from each building, and compares them with TNN and TSNN. Clearly,  observations located at Building 1 benefit much less than those located in the other two buildings from partitioning data into three subsamples according to the predicted Building ID. Recall that Building 1 contains the least number of APs according to Fig. \ref{fig2:third}. Moreover, an additional partition of each subsample into upper-level/lower--level floors does not significantly reduce the mean positioning error.

 \begin{table}[!h]
\begin{center}
\caption{Building-wise mean positioning error (m)}
\begin{tabular}{lcccccccc}
\hline
Building ID      &0 &    1  &  2  & Overall \\ 
(\# of obs.)  & (536) & (307) & (268) & (1111)\\\hline
\ \ \ \ TNN   & 12.65    & 12.52         & 12.04       &12.35\\
\ \ \ \ TSNN &8.18 & 11.97  & 10.61 &9.81\\
\ \ \ \ SCNN &7.94 & 12.07  & 10.44 & 9.68\\
\hline
\end{tabular}
\label{table:3}
\end{center}
\end{table}

Table \ref{table:1} compares our localization results with those provided by the the EvAAL 2015 Competitors in terms of Floor hit rate, Building hit rate, and Mean positioning error. The proposed algorithm achieves 100\% success in estimating the correct building, and an overall performance  95.52\% when considering the floor hit rate. Moreover, the proposed algorithm performs inbetween the winner teams in the EvAAL 2015 competition.



\begin{table}[!h]
\begin{center}
\caption{Localization results on {\it UJIIndoorLoc} validation dataset (1111 observations) vs. four best methods in \cite{b7}.}
\begin{tabular}{lccccc}
& Bldg. hit rate &Floor hit rate &Mean positioning error (m)\\
\hline
SCNN          & 100\%   & 95.52\% & 9.68\\\hline
RTLS\@UM & 100\%   & 93.74\%  &6.20 \\
ICSL            & 100\%   & 86.93\%   &7.67\\
HFTS         & 100\%    & 96.25\%    &8.49\\
MOSAIC    & 98.65\% & 93.86\%   &11.64\\
\hline
\end{tabular}

\label{table:1}
\end{center}
\end{table}

%
%

\section{Conclusion}

In this paper, we proposed a structural ML approach for fingerprinting-based indoor localization problems. We motivated the need for constructing structural components, i.e. locations of APs and shrinkage of location space to a neighborhood area, to improve the accuracy of localization solutions. The use of estimated locations of APs enable us to produce pixel-style RSSI pictures for each observed RSSI vector. The constructed RSSI pictures allow us to use CNN for dealing with the spatial dependence of RSSI readings naturally.  Moreover, the proposed tree-structured binary classification procedure helps us select a subsample from the training dataset for a localization solution, which greatly simplifies the relationship to be learned by the algorithm in the last stage. Therefore, by using the publicly available {\it UJIIndoorLoc} dataset, we show that the proposed structural ML approach can improve the performance of localization solutions. Code for this project can be found at \cite{b17}.







\bibliographystyle{IEEEtran}
\bibliography{overall}

\end{document}